\documentclass[conference]{IEEEtran}
\IEEEoverridecommandlockouts

\usepackage{balance}
\usepackage{helvet}
\usepackage{cite}
\usepackage{amsmath,amssymb,amsfonts}
\usepackage{algorithmic}
\usepackage{graphicx}
\usepackage{subcaption}
\usepackage{textcomp}
\usepackage{xcolor}
\usepackage[hidelinks]{hyperref}
\def\BibTeX{{\rm B\kern-.05em{\sc i\kern-.025em b}\kern-.08em
    T\kern-.1667em\lower.7ex\hbox{E}\kern-.125emX}}

\begin{document}

\title{
A RISC-V Multicore and GPU SoC Platform with a Qualifiable Software Stack for Safety Critical Systems 
}

\author{\IEEEauthorblockN{Marc Sol\'e i Bonet$^{\ast\dagger}$
~~~Jannis Wolf$^{\dagger}$~~~Leonidas Kosmidis$^{\dagger\ast}$
}
\IEEEauthorblockA{
\\
$^\ast$Universitat Polit\`{e}cnica de Catalunya (UPC)
~~~ $^\dagger$Barcelona Supercomputing Center (BSC) \\
}
}

\IEEEaftertitletext{\vspace{-1.0\baselineskip}}

\maketitle

\begin{abstract}
In the context of the Horizon Europe project, METASAT, a hardware platform was developed as a prototype of future space systems.
The platform is based on a multiprocessor NOEL-V, an established space-grade processor, which is integrated with the SPARROW AI accelerator and connected to a GPU, Vortex. Both processing systems follow the RISC-V specification.
This is a novel hardware architecture for the space domain as the use of massive parallel processing units, such as GPUs, is starting to be considered for upcoming space missions due to the increased performance required to future space-related workloads, in particular, related to AI.
However, such solutions are only currently adopted for New Space, since their limitations come not only from the hardware, but also from the software, which needs to be qualified before being deployed on an institutional mission.
For this reason, the METASAT platform is one of the first endeavors towards enabling the use of high performance hardware in a qualifiable environment for safety critical systems. 
The software stack is based on baremetal, RTEMS and the XtratuM hypervisor, providing different options for applications of various degrees of criticality. 

The platform has been tested with space-relevant AI workloads taking full advantage of the hardware resources, even when multiple tasks are sharing the GPU.

\end{abstract}


\section{Introduction}
The continuous advances in software, which are becoming commonly widespread in high-performance environments, have caught the attention of the critical systems domains. 
However, managing such complexity with limited resources and while ensuring some safety in a fast changing environment is a big challenge to tackle.
Nonetheless, the industry and research facilities around the globe are moving towards improving the capabilities of both the existing hardware and software to support highly demanding applications in a safety critical environment.

To that end, model-based approaches are being introduced in the design of space systems.
The Horizon Europe project METASAT~\cite{metasatProject}\cite{metasat_samos}\cite{edhcp2023}, funded by the European Commission, has just completed its work towards the development of model-based design solutions to address the complexity associated with programming advanced high-performance platforms, including AI accelerators and GPUs.

In order to implement and evaluate the METASAT solutions, a high-performance platform for space was required.
However, the current state-of-the-art for space systems is limited to a few low-power processors which are highly constrained due to the difficulties of operating in space.
For that reason, as part of the METASAT effort, a low Technology Readiness Level (TRL 3-4) prototype platform demonstrator of a future on-board computer targeting  was developed.
The complexity of the system is much higher than any existing space processor, as it incorporates an AI accelerator integrated within the processors as well as a GPU. 
The design has been prototyped on a Field-Programmable Gate Array (FPGA) and it is aimed towards satisfying the requirements of the space industry.

Since before the METASAT project there was no space-grade platform with such features, there was also no existing software support for AI and GPU accelerators. 
At the same time, the difficulties of producing qualifiable software for such systems also limited the hardware in use.
To solve this second part of the stalemate, all software used in METASAT is qualifiable or already qualified. 
We have proposed innovative solutions to manage the hardware complexity when using Real-Time Operating Systems (RTOS) and even in the context of a hypervisor,
with multiple partitions accessing the hardware resources at the same time.

The prototype was evaluated with three different use cases with artificial intelligence (AI) workloads relevant for space.
The implementation was done using the TensorFlow Lite Micro framework which we adapted to support the METASAT hardware.
In this scenario we demonstrated the viability of our approach as we were able to execute and accelerate the applications in a qualifiable software environment.
Since the project targeted low TRL, the goal was not performance or reliability, but the proof of the feasibility of our approach, i.e. that high performance accelerators can be used with a qualifiable software stack.

\section{Hardware Platform}

The METASAT platform has been prototyped on the Xilinx Virtex Ultrascale+ VCU118 FPGA, and it is divided in two main blocks, the multicore CPU (Central Processing Unit)  i.e. the \textit{host} processor and a GPU (Graphics Processing Unit), i.e. the \textit{device}.

 The CPU side is a quad-core system based on the NOEL-V processor \cite{noelv}, a RISC-V space grade processor developed by Frontgrade Gaisler. 
 Each of the cores, which are configured to their high-performance configuration, has been extended with SPARROW~\cite{sparrow}, a single instruction multiple data (SIMD) unit for AI acceleration. SPARROW is integrated within the processor integer pipeline, reusing the integer register file to reduce the resource overhead introduced by traditional vector units.

 For the GPU side, we are using Vortex~\cite{vortex}\cite{vortex_micro}, an open-source RISC-V soft-GPU developed at Georgia Tech. 
 Since the original Vortex GPU hardware and its software target x86 host processors, we had to modify the design in order to integrate it with the multicore part of the platform in the same SoC which is emulated on the FPGA in its entirety.
 Since the aim of the platform is to provide a proof of concept we implemented a small configuration of the GPU with a single compute unit, 4 warps and 4 threads.

 Taking advantage of the two separate memory units in the VCU118 each of the subsystems had exclusive access to its own memory, 
 to avoid one system to corrupt each other's memory.
 The communication between the CPU and GPU was established using an AXI-lite connection with the CPU side being the manager.
 We developed our own AXI controller to supervise the GPU execution as well as to handle the CPU-GPU data transmission in both directions.

 A digital twin of the platform was also developed based on QEMU and the connection to the GPU was replicated using Transaction-Level Modeling (TLM) and the SystemC output from Vortex in verilator.

\section{Software Stack}
Albeit all the components of the platform are RISC-V, we require different compilers for the code generated for each of the subsystems.
In particular, for the NOEL-V code we require Gaisler's own variant of GCC, which has been modified to support SPARROW operations.
On the other hand, to compile the kernel code for the GPU we had to use a 32-bit version of the RISC-V GCC, which comes with Vortex.

Although the METASAT platform is Linux-capable, Linux is a complex operating system (OS) which will not be qualified for space~\cite{SMCIT}.
For this reason we had to port the Vortex driver to our platform and adapt it to be used in a baremetal environment.
Since we do not have any file system, this required to embed the precompiled GPU kernel binary into the CPU code using \verb|xxd| to generate a header file that can be included in the source code.

This approach worked as expected with the only limitation 
that OpenCL which is supported in the original Vortex using POCL, cannot be used. The reason is that the OpenCL API relies on the filesystem in order to load the precompiled GPU kernel. Thereforewe decided to work with the Vortex API driver instead.

METASAT 
supports the Real-Time Executive for Multiprocessor Systems (RTEMS) 6 is widely used space systems, and has already been prequalified by the European Space Agency (ESA). 
RTEMS also provides OpenMP support, which enables and simplifies the use of parallelism in the multicore. 
In order to be able to use the GPU under this real-time operating system we just needed to recompile the GPU driver with the RTEMS compiler, while the kernel code is compiled using the Vortex compiler as before.
The modified RTEMS compiler used for the project will be released open-source in the near future.

Finally, the METASAT platform also supports the  XtratuM~\cite{xtratum} hypervisor, which has been developed specifically for space systems and it is space qualified for criticality category B.
Using XtratuM we can have multiple partitions executing simultaneously in different or the same core with the hypervisor managing the context switch.

In order to allow multiple tasks to share the GPU, a feature that Vortex currently lacks, 
we implemented a GPU manager partition, which grants each GPU partition exclusive access to the device. The system uses a queue to handle incoming partitions requests to avoid starvation. 
This requires that before every GPU partition wants to use the GPU it needs to check if it is free, and receive the permission from the GPU manager.
After using it, the GPU partition must notify the manager that the GPU is free again.

While this approach has an overhead in performance as the GPU manager needs to be scheduled before every GPU use, it guarantees the safe access to the resources while allowing multiple partitions to be scheduled.

\section{Use Cases}

In order to evaluate the platform functionality two space-relevant AI applications were ported to support each of the processing units in the platform.
This was done using TensorFlow Lite Micro, an AI framework targeting microcontrollers.
The most relevant layers in TensorFlow where modified to support SPARROW and the Vortex GPU and then the library was compiled with the corresponding compiler, both for baremetal and RTEMS.

\subsection{Cloud screening}
\begin{figure}
    \centering
    \includegraphics[width=1\linewidth]{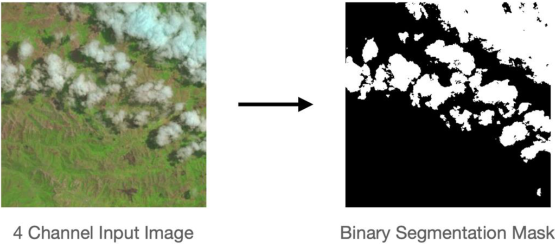}
    \caption{Example of cloud segmentation}
    \label{fig:cloud}
\end{figure}

The first of the applications consists of a cloud screening application.
Cloud screening is the task of identifying and classifying cloud-covered areas in satellite imagery, supporting applications like weather forecasting and climate monitoring.
Figure \ref{fig:cloud} shows an input image from the training set and the output of the model, which is a binary map distinguishing between pixels that represent a cloud (1, white) and pixels that do not (0, black).

For the cloud screening, a small fully convolutional U-Net was trained on the Cloud95 dataset. 
The model was designed to classify cloud pixels in satellite images, effectively distinguishing between cloud-covered and clear areas.

Noth applications follow the same structure, 
and consist of three main modules that work together to process and generate the results:

\begin{itemize}
    \item \textbf{Image Loader}: The module is responsible for loading the image data, which is embedded in the application, in the same way as the Vortex sGPU kernel binary. 
    \item \textbf{AI Inference Module}: the core of the AI application, which performs the inference on the loaded image using Tensor Flow Lite Micro. 
    The inference can be performed on one of the three backends that were developed for the METASAT platform: CPU, SPARROW and Vortex GPU.
    \item \textbf{Output Handler}: This module represents the output of the model to the screen, in the cloud screening this is the percentage of cloud covered area in the input image.
\end{itemize}

\subsection{Ship detection}

\begin{figure}
    \centering
    \includegraphics[width=1\linewidth]{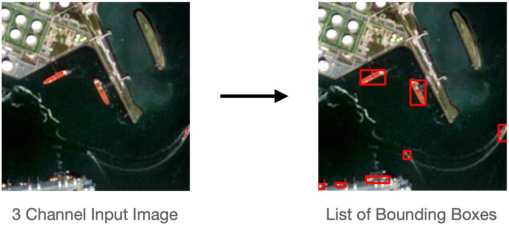}
    \caption{Example of ship detection}
    \label{fig:ship}
\end{figure}

The second use case is a ship detection application, which refers to the task of identifying and localizing ships in satellite imagery, aiding in maritime surveillance, logistics and environmental monitoring.
Figure \ref{fig:ship} shows on the left the input to the model and on the right the output, which is a list of bounding boxes around ships that were detected on the image.

For the ship detection, a YOLOX-Tiny model was trained on a subset of the large Airbus Ship Detection dataset~\cite{airbus-dataset}.
This dataset consists of annotated satellite images containing ships in various environments and conditions.
YOLOX-Tiny, a lightweight variant of the YOLOX architecture, was chosen for its balance of speed and accuracy in detecting objects in images.

In this test the output corresponds to a list of the boxes surrounding found ships.
The output from the model varies slightly from the ground truth of the test data, but is with an intersection over union of 92.6\%, which is perfectly acceptable.

Both cloud screening and ship detection are part of the ESA open source benchmarking suite OBPMark-ML~\cite{OBPMark}\cite{OBPMark-OBDP}, 
which focuses on machine learning (ML) related workloads.

\section{Evaluation}

The evaluation of the METASAT platform was performed using the uses cases mentioned before.
The code has been loaded to the VCU118 FPGA using GRMON, a tool provided by Gaisler which allows to manage the connection with a NOEL-V system on an FPGA.

The first initial tests were testing each of the use cases in isolation in baremetal, RTEMS and RTEMS when running in a partition within the hypervisor.
The results for the cloud screening and ship detection can be seen in Tables \ref{tab:cloud} and \ref{tab:ship} respectively.

\begin{table}[h]
    \centering
    \caption{Cloud screening execution times in baremetal, RTEMS and XtratuM when using different hardware resources}
    \label{tab:cloud}
    \begin{tabular}{|l|l|l|l|}
    \hline
    \textbf{Accelerator} & \textbf{Baremetal} & \textbf{RTEMS} &  \textbf{XtratuM} \\
    \hline
    None (CPU) & 33.359 s & 33.215 s & 32.664 s \\
    \hline
    SPARROW & 16.873 s & 19.546 s & 17.422 s \\
    \hline
    Vortex (GPU) & 58.800 s & 59.576 s & 58.561 s \\
    \hline
    \end{tabular}
\vspace{-1em}
\end{table}
\begin{table}[h]
    \centering
        \caption{Ship detection execution times in baremetal, RTEMS and XtratuM when using different hardware resources}
    \label{tab:ship}
    \begin{tabular}{|l|l|l|l|}
    \hline
    \textbf{Accelerator} & \textbf{Baremetal} & \textbf{RTEMS} &  \textbf{XtratuM} \\
    \hline
    None (CPU) & 671.647 s & 675.730 s & 664.754 s \\
    \hline
    SPARROW & 265.909 s & 290.457 s & 290.004 s \\
    \hline
    Vortex (GPU) & 786.022 s & 786.095 s & 784.868 s \\
    \hline
    \end{tabular}
\end{table}

The first thing noticeable from the results is a loss of performance when using the GPU compared to both SPARROW and the CPU.
This can be explained due to the difference in the two systems' configuration. 
While the NOEL-V CPU is configured in its high-performance mode, with 64-bits, L2-cache and dual-issue; Vortex is configured as single 32-bit core with 4 warps and 4 threads.
We also need to take into account the communication overhead when sending the data from the CPU to GPU and vice-versa.

The goal of the METASAT project was not to achieve high performance but evaluate the techniques and be able to generate software for different types of accelerators.
In our previous work with COTS GPUs~\cite{date_gpu4s} we have demonstrated that the speed-up offered by GPUs compared to CPU processing 
is considerable and justifies this approach. We are confident that on a larger implementation with more resources on the GPU we could achieve similar speed-ups.

\begin{figure}
\centering
\begin{subfigure}{.25\textwidth}
    \centering
    \includegraphics[width=0.95\linewidth]{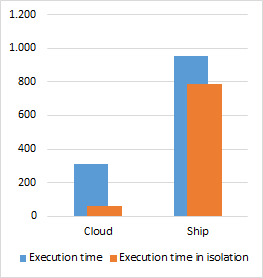}
    \caption{GPU and GPU}
    \label{fig:gpu-gpu}
\vspace{1em}
\end{subfigure}%
\begin{subfigure}{.25\textwidth}
    \centering
    \includegraphics[width=0.95\linewidth]{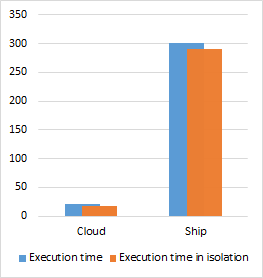}
    \caption{SPARROW and SPARROW}
    \label{fig:sparrow-sparrow}
\vspace{1em}
\end{subfigure} %
\begin{subfigure}{.25\textwidth}
    \centering
    \includegraphics[width=0.95\linewidth]{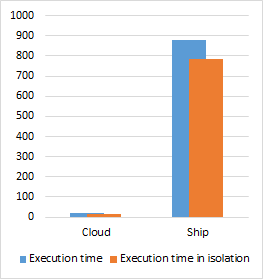}
    \caption{SPARROW and GPU}
    \label{fig:sparrow-gpu}
\end{subfigure}%
\begin{subfigure}{.25\textwidth}
    \centering
    \includegraphics[width=0.95\linewidth]{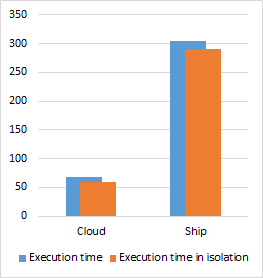}
    \caption{GPU and SPARROW}
    \label{fig:gpu-sparrow}
\end{subfigure}
    \caption{Execution time of two XtratuM partitions with cloud screening and ship detection using SPARROW and the GPU and compared with each partition execution time in isolation}
    \label{fig:cloud-ship}
    \vspace{-1em}
\end{figure}

In order to evaluate the behavior of the platform when executing two simultaneous partitions in different cores, we designed a second set of experiments where a partition running the cloud screening and another running the ship detection would be executed at the same time.
In Figure~\ref{fig:cloud-ship}, the results of the experiments are shown when running the partitions in all possible combinations of SPARROW and GPU and compared to when running in isolation.

In general the overhead of running the two partitions simultaneously is relatively small (1.05$\times$ - 1.2$\times$), the only exception being the GPU-GPU case where the slowdown is over 5$\times$ for the cloud screening partition.
This can be attributed to the sharing of the GPU where the execution cannot continue until the previous partition has released the device.
This results are similar to what we have seen when time sharing a GPU in a commercial-off-the-shelf (COTS) device.

The same applications were also executed on the digital twin.
However, the long execution time of the ship detection combined with the slow simulation of the TLM SystemC interface made it impossible to complete the execution of the GPU simulation for the ship detection.
For this reason, only the cloud detection results are presented in Table~\ref{tab:cloud-twin}.
These results correspond to the XtratuM variant of the test and the timings shown correspond to the execution time in the FPGA, the simulated time which is the time reported by RTEMS 
and the simulation time which is how long took for the digital twin to complete the simulation.

The more impactful result is the slow GPU simulation time in contrast with the reported simulated time, which is better than for the CPU.
These results show that the QEMU simulation is quite accurate in terms of the simulated execution time and the simulation is fast enough to provide a relevant result.
On the other hand, the simulation of the GPU using the TLM and SystemC approach is too slow, and, while it is viable, it is a future optimization point.

\begin{table}[h]
    \centering
        \caption{Cloud detection simulation time in the digital twin}
    \label{tab:cloud-twin}
    \begin{tabular}{|l|l|l|l|}
    \hline
    \textbf{Accelerator} & \textbf{Execution time} & \textbf{Simulated time} &  \textbf{Simulation time} \\
    \hline
    None (CPU) & 32.664 s & 30.136 s & 56 s \\
    \hline
    SPARROW & 17.423 s & 11.586 s & 25 s \\
    \hline
    Vortex (GPU) & 58.561 s & 20.039 s & 90924 s \\
    \hline
    \end{tabular}
\end{table}

\section{Conclusions}
The METASAT project has explored the use of the complex parallel hardware in a safety critical environment when using a system with a multicore, extended with an AI accelerator, and a GPU.
The integration of the two subsystems has been done successfully and the approach has been demonstrated in different space relevant workloads.

The platform works with software which is either qualified for space or can be qualified in the future. 
This includes the compilers as well as the hypervisor layer, which allows the execution of simultaneous applications sharing the hardware resources with minimal overhead.
The METASAT platform will be released as open-source in the near future and will be available at~\cite{metasatGitLab}.

\vspace{-0.2em}
\section*{Acknowledgements}

This work was supported by the European Community’s Horizon Europe programme under the METASAT project (grant agreement 101082622). In addition, it was partially supported by the Spanish Ministry of Economy and Competitiveness under the grant 
IJC-2020-045931-I ( Spanish State Research Agency / Agencia Espa\~nola de Investigaci\'on (AEI) / http://dx.doi.org/10.13039/501100011033 ) and by the Department of Research and Universities of the Government of Catalonia with a grant to the CAOS Research Group (Code: 2021 SGR 00637).
\vspace{-0.5em}

\bibliographystyle{plain}
\bibliography{biblio}

\end{document}